\documentclass[10pt]{iopart}

\usepackage{amsfonts}
\usepackage{theorem,verbatim}
\usepackage[english]{babel}
\usepackage[ansinew]{inputenc}
\usepackage[T1]{fontenc}
\usepackage{color}
\usepackage{graphicx}
\usepackage{wasysym}
\usepackage{rotating}

\newcommand{\be}{\begin{equation}}
\newcommand{\ee}{\end{equation}}
\newcommand{\dd}{\mathrm{d}}
\newcommand{\trans}{^\mathrm{T}}

\newcommand{\R}{\mathbb{R}}

\newcommand{\vectornorm}[1]{\vert\vert#1\vert\vert}

\begin{document}

\title[Reduced-order model for EIT based on POD]{Reduced-order model for
electrical impedance tomography based on proper orthogonal decomposition}
\author{A~Lipponen$^1$, A~Sepp{\"a}nen$^1$ and J~P~Kaipio$^{1,2}$}
\address{$^1$ Department of Applied Physics, University of Eastern Finland,
FI-70211 Kuopio, Finland}
\address{$^2$ Department of Mathematics, University of Auckland, Auckland 1142,
New Zealand}
\ead{Antti.Lipponen@uef.fi}

\begin{abstract}
Electrical impedance tomography (EIT) is an imaging modality in which the conductivity
distribution inside a target is reconstructed based on voltage measurements from
the surface of the target.
Reconstructing the conductivity distribution is known to be an ill-posed inverse problem,
the solutions of which are highly intolerant to modelling errors.
In order to achieve sufficient accuracy,
very dense meshes are usually needed in a finite element
approximation of the EIT forward model.
This leads to very high-dimensional problems and often
unacceptably tedious computations for real-time applications.
In this paper, the model reduction in EIT is considered
within the Bayesian inversion framework.
We construct the reduced-order model
by proper orthogonal decompositions (POD)
of the electrical conductivity and the potential distributions.
The associated POD modes are computed based on
{\it a priori} information
on the conductivity.
The feasibility of the reduced-order model is tested both numerically
and with experimental data.
In the selected test cases, the proposed model reduction approach
speeds up the computation by more than two orders of magnitude
in comparison with the conventional EIT reconstruction,
without decreasing the quality of the reconstructed images significantly.
\end{abstract}



\submitto{\IP}

\maketitle

\vspace{-6pt}


\section{INTRODUCTION}
\vspace{-2pt}

In electrical impedance tomography (EIT), a set of alternating
currents are injected to the target through an array of boundary electrodes, 
and the resulting voltages on the electrodes are measured.
Based on these measurements, the internal conductivity distribution 
of the target is reconstructed.
The applications of EIT include e.g.
industrial process monitoring, 
\cite{hosseini2010,pakzad2008,park2008,takriff2009},
control \cite{gutierrez2000} and design \cite{williams1997},
medical imaging \cite{frerichs2000,zou2003}
and geophysical exploration
\cite{brunet2010,gokturkler2008,perrone2004}.
Advantages of EIT are the very good temporal resolution 
(up to 1000 frames per second \cite{williams1995b}),
the non-invasive and radiation-free nature, 
and affordable measuring devices.
A drawback of EIT is often the relatively low spatial resolution,
which is partly due to the diffusive nature of the modality
and partly a result of simplifications that are typically made in
image reconstruction in order to compute the reconstructions in
the required time frame.

The image reconstruction problem in EIT is known to be 
an ill-posed inverse problem, i.e. the
solutions are very sensitive to measurement noise and modelling errors.
Therefore, an accurate model for the measurements 
-- i.e. the forward model --
is needed in order to obtain feasible reconstructions.
The most accurate forward model
for a physically realizable
EIT measurements is referred to as 
the complete electrode model (CEM) \cite{cheng1989}.
The CEM consists of an elliptic partial differential equation (PDE) 
and associated boundary conditions.
The CEM is usually approximated with the finite element method (FEM) where
finite dimensional approximations to the electrical conductivity and potential are
written using locally supported piecewise polynomial basis functions.
Typically, dense finite element (FE) meshes are needed in order to model the
measurements with sufficient accuracy.
Therefore, the use of locally supported FE-bases often leads to a high dimensional 
problem which may require a large amount of memory and time to solve.
The image reconstruction problem of EIT is a non-linear inverse problem,
and in optimization based reconstruction methods,
the conductivity estimate is computed iteratively.
In the case of a conventional FE approximation of CEM,
the large dimensional forward problem
needs to be solved in each iteration step.
However, in many industrial and medical applications 
online reconstructions are needed, and
the allowable time for the computations is short -- 
sometimes even in order of milliseconds.
Usually, the problem of high computational demand is overcome by using 
coarse FE-meshes \cite{kaipio2005,nissinen2009} and/or 
global linearization of the observation model \cite{vauhkonen1999a,pursiainen2006b}
or even highly simplified back projection reconstruction methods \cite{barber1984,santosa1990}.
Such choices, however, often lead to severely biased reconstructions.
It is thus of great importance to develop new methods for 
model reduction in EIT,
so that the computation time could be decreased
without ruining the quality of the reconstructions.

The proper orthogonal decomposition (POD) \cite{jolliffe2002}
has been widely used in different types of data and model reduction problems.
In data reduction, 
data vectors are represented in a low dimensional space spanned by POD modes
which are eigenvectors of the covariance matrix of the data set.
The data reduction applications of the POD include e.g.
image processing \cite{zhang2009}, face recognition 
\cite{draper2003,kirby1990,sirovich1987b},
and data clustering \cite{yeung2001}.
Further,
the POD has been applied to the reduction of PDE models
\cite{cao2006,deane1991,holmes1998,ma2002,ravindran2000}.
In these papers, the basis functions for the FE approximation of the PDE were
constructed by using the POD.
The POD modes were computed based on the conventional FE approximation of the
PDE with some fixed parameters.

The POD and other projection-based model reduction methods
have also been applied to inverse problems spanned by PDEs.
Firstly, the POD has been applied to 
constraining the solution of the inverse problem
(i.e. the unknown parameter distribution)
to a desired subspace
which is selected based on the known properties of the target.
In EIT, the POD modes for the internal conductivity distribution
have been selected based on anatomical information in imaging of 
thorax \cite{vauhkonen1997}.
Respectively, in \cite{banks2000,banks2003} the POD 
was used to construct a reduced-order basis for the unknown material parameters
in other electromagnetic inverse problems.
Secondly,
the POD has been used for reducing the number of basis functions
for the state of the system (i.e. the solution of the PDE).
For example,
in \cite{jin2008} the temperature distribution was represented in a
reduced-order POD basis in an inverse heat conduction problem.
Here, the inverse problem was to estimate a scalar Robin coefficient
using temperature measurements from the boundary
of the domain.
An ensemble of temperatures for constructing the POD basis 
was obtained by solving the heat equation corresponding 
to an {\it ad hoc} selected set of
Robin coefficients.
For other studies where
the POD modes were used for constructing the FE basis for the
solution of a PDE, see
\cite{cao2006,deane1991,holmes1998,ma2002,ravindran2000}.
Note that in all these papers,
the number of unknown parameters in the inverse problem
was small, and the POD was applied {\it only} in
the state of the system.
Recently, Lieberman et. al. \cite{lieberman2010} proposed an approach to
the model reduction for high-dimensional statistical inverse problems.
They applied a heuristic optimization-based greedy sampling
for simultanously constructing the bases for {\it both} 
the unknown high dimensional parameter distribution 
{\it and} the state of the system.
The feasibility of the 
proposed approach was demonstrated with an
example of a
statistical inverse problem
of groundwater flow.
By using the reduced model, 
they computed an estimate for the posterior distribution of the 
hydraulic conductivity parameters by 
a Markov Chain Monte Carlo (MCMC) method.

In this paper,
we propose a POD-based reduced-order model for EIT.
As in \cite{lieberman2010},
we represent both the parameters (the electrical conductivity distribution) and
the state of the system (the electrical potential distribution)
in reduced bases.
We consider the model reduction within the Bayesian framework,
and write an explicit statistical prior model for the unknown conductivity distribution.
Unlike in \cite{lieberman2010},
we construct the reduced order basis for the conductivity
based on the prior distribution.
Moreover, we compute the realizations of the potential field
using samples drawn from the prior distribution of the conductivity,
and use the ensemble of potential fields to
construct the POD basis for the potential.
Finally, 
we complete the reduced model by approximating
the errors related to the reduced-order approximations
by an auxiliary additive noise process.
In the last step,
the so-called approximation
error approach \cite{kaipio2007a,kaipio2005}
is adopted.

The rest of the paper is organized as follows.
In Section \ref{sect.eit}, the EIT imaging is briefly reviewed;
the discussion is limited to the computational aspects of the forward model 
and the Bayesian inversion.
The POD based model reduction for EIT is proposed
in Section \ref{sect.pod}.
Further, in Section \ref{sect.examples}, the reduced-order reconstruction is 
evaluated both with simulated and experimental data.
Finally, the conclusions are drawn in Section \ref{sect.conclusions}.

\vspace{-6pt}

\section{ELECTRICAL IMPEDANCE TOMOGRAPHY} \label{sect.eit}
\vspace{-2pt}

In this section, computational aspects of EIT imaging are considered
in Bayesian (statistical) framework.
The complete electrode model (CEM) and its
FE approximation are
briefly reviewed in Sections \ref{subsec.CEM} and \ref{subsec.FEMCEM},
respectively.
The Bayesian inverse problem of EIT is discussed in
Section \ref{subsec.EITbayes}.


\subsection{Complete electrode model}
\label{subsec.CEM}

In EIT, the most accurate measurement model is the CEM \cite{cheng1989},
which consists of the PDE
\be
\nabla\cdot(\sigma\nabla u)=0,\ \ \ \ \ {\vec r}\in \Omega \label{eq.CEM1}
\ee
and the following boundary conditions:
\begin{eqnarray}
u+z_{\ell}\sigma \frac{\partial{u}}{\partial {\vec n}} &=& U_{\ell},\ \ \ {\vec r}\in
e_{\ell},\ \ell=1,2,\ldots,L \label{eq.CEM2}\\
\sigma \frac{\partial{u}}{\partial {\vec n}} &=& 0,\ \ \ \ {\vec r}\in \partial\Omega
\backslash \cup_{\ell=1}^L e_{\ell}\label{eq.CEM3}\\
\int_{e_{\ell}}\sigma\frac{\partial{u}}{\partial {\vec n}}\, \mathrm{d}S & = & I_{\ell},
\ \ \ \ell=1,2,\ldots,L\label{eq.CEM4}
\end{eqnarray}
where $\sigma = \sigma(\vec{r})$ is the electrical conductivity, $u = u(\vec{r})$ is 
the electric potential inside the target domain $\Omega$, 
also referred to as the inner potential,
${\vec r}$ is the spatial coordinate
and ${\vec n}$ is the unit outward normal vector.
Contact impedances, electrode potentials, and injected currents 
corresponding to the electrodes
$e_\ell,\ \ell=1,\ldots,L$ are denoted by $z_\ell$, $U_\ell$, and $I_\ell$, respectively.
For further use, we denote the vectors 
$z=[z_1,\ldots,z_L]^{\mathrm T},\ U=[U_1,\ldots,U_L]^{\mathrm T}$ and
$I=[I_1,\ldots,I_L]^{\mathrm T}$.
In addition to (\ref{eq.CEM1}--\ref{eq.CEM4}), we write
\be
\sum_{\ell=1}^L I_\ell = 0,\qquad \sum_{\ell=1}^L U_\ell = 0. \label{eq.CEM5}
\ee
where the former condition ensures that the charge conservation law is fulfilled
and the latter one fixes the reference potential level.

The forward problem of EIT is to solve the inner potential $u = u(\vec{r})$
and the electrode potentials $U_\ell,\ \ell=1,\ldots,L$, given the
conductivity distribution $\sigma(\vec{r})$, the contact impedances $z_\ell$ 
and the injected currents $I_\ell$.
The existence and uniqueness 
of the solution was proven in \cite{somersalo1992}.
The measurements in EIT imaging consist of potential differences, 
i.e. voltages, between electrodes.
The inverse problem in EIT is to reconstruct the 
conductivity distribution $\sigma(\vec{r})$ given the voltage measurements
corresponding to various sets of electrode currents.


\subsection{Finite element approximation of the CEM}
\label{subsec.FEMCEM}

The variational form of (\ref{eq.CEM1}--\ref{eq.CEM4})
can be written as \cite{somersalo1992}
\begin{equation}
B((u, U), (v, V))=\sum_{l=1}^L I_l V_l \ , \ \ \forall (v, V)\in H
\label{eq.variational.form}
\end{equation}
where $H=H^1(\Omega) \times {\mathbb{R}}^L$,
$H^1(\Omega)$ is a Sobolev space
and
$B:H \times H \rightarrow {\mathbb{R}}$ is a bilinear form such that 
\be
B((u, U), (v, V))=\int_{\Omega} \sigma \nabla u \cdot \nabla v
\mathrm{d}x
+\sum_{l=1}^L\frac{1}{z_l}\int_{e_l}(u-U_l)(v-V_l)\mathrm{d}S
\label{eq.var2}
\ee
In this section, the FE approximation is reviewed briefly.
For details, see \cite{vauhkonen1999a}.
Later, in Section \ref{sect.pod}, we point out how the reduced order 
approximations affect the FE scheme.

In a finite element implementation of the CEM, the finite dimensional
approximations for 
the conductivity
and the inner potential $u$ 
are written as
\be
\sigma \approx \sigma^h =\sum_{i=1}^{N} \alpha_i\phi_i(\vec r),
\ \ \ \ \ 
u \approx u^h = \sum_{j=1}^M \beta_j \psi_j(\vec r).
\label{eqFEappr}
\ee
Here 
$\sigma^h=\sigma^h(\vec{r})$
denote the approximated conductivity
and inner potential
respectively,
$\phi_i = \phi_i(\vec{r}),\ i=1,2,\ldots,N$ 
and
$\psi_j = \psi_j(\vec{r}),\ j=1,2,\ldots,M$ 
are the basis functions, and
$\alpha = \left[\alpha_1,\alpha_2,\ldots,\alpha_M\right]\trans$,
$\beta = \left[\beta_1,\beta_2,\ldots,\beta_M\right]\trans$ 
are the
corresponding coefficients.
In the discussion below, we identify $\sigma$ and its finite
dimensional representation $\alpha$.
Usually, the basis functions $\phi_i$ and $\psi_i$ are selected as piecewise linear or
piecewise quadratic functions.
For higher order polynomial approximations, see \cite{pursiainen2006a}.
The potentials $U$ on the electrodes are written as
\be
U = \sum_{k=1}^{L-1} \gamma_k n_k \label{eq.U}
\ee
where $n_k$ are the basis functions chosen as
$n_1 = [1,-1,0,\ldots,0]\trans$, $n_2 = [1,0,-1,\ldots,0]\trans$, $\ldots$, 
$n_{L-1} = [1,0,0,\ldots,-1]\trans$ to fulfill the latter condition in (\ref{eq.CEM5}).

Inserting the above approximations to the variational form
(\ref{eq.variational.form})
leads to the following matrix equation \cite{vauhkonen1999a}
\be \label{eq.atf}
A \mathbf{\theta} = f
\ee
where 
$\mathbf{\theta} = [\beta\trans,\ \gamma\trans]\trans \in \R^{(M+L-1) \times 1}$,
$\gamma = [\gamma_1,\ldots,\gamma_{L-1}]\trans$ 
and the vector
$f \in \R^{(M+L-1) \times 1}$ is defined as
\be
f = \left[ \begin{array}{c} \mathbf{0}_{M} \\ 
\mathcal{C}\trans I \end{array} \right].
\ee
Here, $\mathbf{0}_{M} \in \R^{M \times 1}$ is a vector of zeros
and
$\mathcal{C} = \left[n_1,\ldots,n_{L-1}\right] \in \R^{L \times (L-1)}$.
Furthermore, matrix $A = A(\sigma,z) \in \R^{(M+L-1)\times (M+L-1)}$ is of the form
\be
A = \left[ \begin{array}{cc} B + D & E \mathcal{C} \\ \mathcal{C}\trans E\trans & \mathcal{C}\trans F \mathcal{C} \end{array} \right]
\ee
where
\begin{eqnarray}
B(i,j) &=&  \sum_{i=1}^{N} \alpha_i \int_\Omega\phi_i\nabla \psi_i \cdot \nabla \psi_j \dd \Omega,\quad 1 \leq i,j \leq M \label{eq.B}\\
D(i,j) &=& \sum_{\ell=1}^L \frac{1}{z_\ell} \int_{e_\ell} \psi_i \psi_j \dd S,\quad 1 \leq i,j \leq M \label{eq.D}\\
E(i,j) &=& -\frac{1}{z_j} \int_{e_j} \psi_i \dd S,\quad 1\leq i \leq M,\ 1\leq j \leq L\\
F(i,j) &=& \sum_{\ell=1}^L \frac{1}{z_\ell} \int_{e_\ell} (n_i)_\ell (n_j)_\ell \dd S = \left\{ \begin{array}{ll} 0,& i\neq j \\ \frac{|e_j|}{z_j},& i=j \end{array} \right. 1 \leq i,j \leq L
\end{eqnarray}
where $|e_j|$ is the measure of the electrode $e_j$, i.e. the length of the electrode
in two dimensions (2D) and the area of the electrode in three dimensions (3D).
The integrals are usually computed numerically 
for example with Gaussian quadratures.

Formally, solution of the system (\ref{eq.atf}) is $\theta = A^{-1}f$.
In practice, however, the system of equations is solved, for example,
with the LU-decomposition.
As noted above, the solution $\theta$ consists of the parameter vectors
$\beta$ and $\gamma$ corresponding to approximations of
internal potential field $u$ and the electrode potentials $U$, respectively.
This is the solution of the forward problem.
It should be noted that if the 
basis functions $\psi_j$ for the internal potential
are locally supported, matrices
$B$ and $D$ are sparse.
Further, the demand for the high accuracy approximations often necessitates the use
dense FE meshes which implies that $B$ and $D$ are high-dimensional.
On the contrary, if the potential $u$ is approximated with a low number of globally
supported basis functions
-- choice made in Section \ref{sect.pod} --
the matrices $B$ and $D$ become low-dimensional and dense.

Based on the finite element approximation described above,
a set of voltages $V$ between selected electrodes can be written in the form
$V=\mathcal{M} \mathcal{C} \gamma$,
where $\gamma=\gamma(\sigma,z,I)$ 
is obtained from the solution of (\ref{eq.atf}),
and 
$\mathcal{M}$ is a difference matrix
referred to as the measurement pattern.
Further, since the dependence of the current pattern $I$ and $\gamma$ is linear,
we can write
\be\label{eq.VRI}
V = R(\sigma,z)I
\ee
where $R=R(\sigma,z)$ is referred to as the resistance matrix, and 
$R(\sigma,z)I=\mathcal{M} \mathcal{C} \gamma$.
In EIT, the voltages $V^{(i)}$ are measured corresponding to several current patterns
$I^{(i)},\ i=1,\ldots,N_{\rm inj}$, where $N_{\rm inj}$ denotes the
number of current patterns.
If the measurement pattern $\mathcal{M}$ is the same for all current injections,
equation (\ref{eq.VRI}) yields
$[V^{(1)},\ldots,V^{(N_{\rm inj})}] = R(\sigma,z)[I^{(1)},\ldots,I^{(N_{\rm inj})}]$.
For the notational convenience, we assume here that the contact impedances $z$ are known;
however, the extension to cases of unknown $z$ is straightforward, see
\cite{vilhunen2002a}.
Then, by stacking all the voltage measurements in one column vector 
$V_{\epsilon} = [(V^{(1)})\trans,\ldots,(V^{(N_{\rm inj})})\trans]\trans$,
we can write the observation model of EIT
\be\label{obs.eq}
V_{\epsilon} = \mathcal{V}(\sigma) + \epsilon
\ee
where 
$\mathcal{V}(\sigma)=[(R(\sigma,z)I^{(1)})\trans,\ldots,
(R(\sigma,z)I^{(N_{\rm inj})})\trans]\trans$
and $\epsilon$ is a stochastic term consisting of the measurement noise.
Note that the observation model (\ref{obs.eq}) is
non-linear with respect to the conductivity distribution $\sigma$.


\subsection{Bayesian inversion in EIT}
\label{subsec.EITbayes}

The solution of the statistical inverse problem in EIT is the
posterior distribution, i.e. the conditional distribution
of $\sigma$ given the voltage measurements.
We denote the associated posterior density by $\pi(\sigma|V_{\epsilon}^\mathrm{meas})$
where $V_{\epsilon}^\mathrm{meas}$ is the realization of the voltages
$V_{\epsilon}$.
Using the Bayes formula, the posterior density can be written in the form
\be
\pi(\sigma|V_{\epsilon}^\mathrm{meas}) = \frac{\pi(V_{\epsilon}^\mathrm{meas}|\sigma) 
\pi_\sigma(\sigma)}{\pi_{V_{\epsilon}}(V_{\epsilon}^\mathrm{meas})}
\propto \pi(V_{\epsilon}^\mathrm{meas}|\sigma) \pi_\sigma(\sigma)
\ee
where $\pi(V_{\epsilon}^\mathrm{meas}|\sigma)$ is the likelihood density
defined by the observation model (\ref{obs.eq}),
and 
$\pi_{V_{\epsilon}}(V_{\epsilon}^\mathrm{meas}) 
= \int\pi(\sigma,V_{\epsilon}^\mathrm{meas}){\rm d}\sigma$
acts as a normalization factor.
Further, $\pi_\sigma(\sigma)$ denotes the marginal density of $\sigma$,
also referred to as the {\it prior density} because it 
includes the information we have on the
conductivity before the measurements.

A practical estimate for the conductivity is obtained by computing some point or
spread estimate from the posterior distribution.
One of the most commonly used point estimate is the {\it maximum a posteriori} 
(MAP) estimate
\be\label{eqMAP}
\sigma_{\rm MAP} = {\rm arg}\max_{\sigma}\pi(\sigma|V_{\epsilon}^\mathrm{meas}).
\ee
The problem of finding the MAP estimate requires solving an optimization problem.
For example,
if $\sigma$ and $\epsilon$ are modeled as 
mutually uncorrelated Gaussian random variables
$\sigma \sim \mathcal{N}(\bar{\sigma}, \Gamma_{\sigma})$ and
$\epsilon \sim \mathcal{N}(\bar \epsilon, \Gamma_\epsilon)$,
the MAP estimate gets the form 
\be
\sigma_{\rm MAP} = \arg \min_\sigma \left\{\vectornorm{L_{\sigma} (\sigma - \bar{\sigma})}^2
+ \vectornorm{L_\epsilon (V_{\epsilon}^\mathrm{meas} - \mathcal{V}(\sigma) - \bar \epsilon)}^2 \right\} 
\label{eq.mapmin}
\ee
where $L_\sigma$ and $L_\epsilon$ are the Cholesky factors of 
the inverted covariance matrices $\Gamma_\sigma^{-1}$
and $\Gamma_\epsilon^{-1}$, respectively
\cite{kaipio1999c}.
Due to non-linearity of the mapping $\mathcal{V}(\sigma)$,
the solution of the minimization problem (\ref{eq.mapmin}) is sought 
iteratively by some optimization method, such as the Gauss-Newton method.
In each iteration step, the function $\mathcal{V}(\sigma)$ 
and the Jacobian matrix $J=\frac{\partial\mathcal{V}(\sigma)}{\partial\sigma}$
are evaluated.
For computation of the Jacobian matrix with the so-called adjoint method,
see \cite{kaipio2000}.

It is worth to notice that since the conductivity distribution $\sigma$ is 
a random variable, 
the internal potential distribution $u$ is also stochastic.
This realization is the core of the next section
in which both the conductivity and the potential distribution are represented in
reduced-order bases selected with the POD.


\section{REDUCED-ORDER MODELLING IN EIT BASED ON PROPER ORTHOGONAL DECOMPOSITION} 
\label{sect.pod}
\vspace{-2pt}

In the POD,
samples of (possibly) correlated random variables are converted into
a set of uncorrelated random variables.
In this procedure the original data is projected into a subspace spanned by
orthogonal basis functions referred to as the POD modes.
Depending on the application, 
POD is also known as principal component analysis (PCA)
\cite{jolliffe2002}, Karhunen-Lo\'{e}ve decomposition 
\cite{karhunen1946,loeve1946,loeve1955},
or Hotelling transform \cite{hotelling1933}.
The governing idea of POD is that in a case of highly correlated data, 
a low number of POD modes account for most of the data.
For this reason, POD is often useful for data and model reduction problems.

In the POD based reduced-order model 
the FE approximations  $\sigma^h$ and $u^h$
(see equation (\ref{eqFEappr})) for the conductivity 
and potential are further approximated as
\begin{eqnarray}
\sigma & \approx  \sigma^h \approx
\sigma_0 +
\sum_{i=1}^{\hat N} \alpha^{\mathrm{POD}}_i\ \phi^{\mathrm{POD}}_i (\vec r)
\label{eq.sigmaapprx}\\
u & \approx  u^h \approx
u_0 +
\sum_{j=1}^{\hat M} \beta^{\mathrm{POD}}_j\ \psi^{\mathrm{POD}}_j(\vec r)
\label{eq.uapprx}
\end{eqnarray}
where
$\sigma_0={\mathbb E}\{\sigma\}$ 
and $u_0={\mathbb E}\{u\}$ 
are the expectations of $\sigma$ and $u$,
respectively.
Further,
$\phi^{\mathrm{POD}}_i$ 
and $\psi^{\mathrm{POD}}_i$
are the basis functions
of $\sigma$ and $u$,
and $\alpha^{\mathrm{POD}}_i$ and $\beta^{\mathrm{POD}}_j$ are
the corresponding coefficients.
The basis functions 
$\phi^{\mathrm{POD}}_i$ and $\psi^{\mathrm{POD}}_i$
are selected such that 
${\mathbb E}
\{\Vert \sigma^h -\sigma_0
- \sum_{i=1}^{\hat N} \alpha^{\mathrm{POD}}_i\ \phi^{\mathrm{POD}}_i
\Vert^2 \}$
and
${\mathbb E}\{\Vert u^h - u_0
-\sum_{j=1}^{\hat M} \beta^{\mathrm{POD}}_j\ \psi^{\mathrm{POD}}_j
\Vert^2\} $,
are minimized over all $\hat N$ and $\hat M$ dimensional bases,
respectively.
Such a representation is referred to as the POD
\cite{jolliffe2002}.

In practice, the POD bases $\{\phi^{\mathrm{POD}}_i\}$ and
$\{\psi^{\mathrm{POD}}_j\}$ of dimensions 
$ \hat N$ and $\hat M$
are obtained as eigenvectors 
of conductivity and potential covariances
$\Gamma_{\sigma}$ and $\Gamma_{u}$
corresponding to $ \hat N$ and $\hat M$
largest eigenvalues, respectively.
Here, the covariance matrix $\Gamma_{\sigma}$
is determined by the prior model $\pi_\sigma(\sigma)$,
and the covariance of the electric potential, $\Gamma_{u}$,
is approximated by the sample covariance
\be
\Gamma_u\approx
\frac{1}{T-1}\sum_{i=1}^T(u^{(i)}-u_0)(u^{(i)}-u_0)\trans
\ee
where $u^{(i)}$ is a sample from the potential distribution
and $u_0=\frac{1}{T}\sum_{i=1}^T u^{(i)}$ is the sample mean.
The samples $u^{(i)}$ are computed by 
solving the forward problem 
corresponding to samples $\left\{\sigma^{(i)}\right\}_{i=1}^T$
drawn from the prior distribution of $\sigma$.
That is, 
corresponding to each conductivity sample $\sigma^{(i)}$,
the CEM (\ref{eq.CEM1}--\ref{eq.CEM4})
is approximated with the FEM as
described in Section \ref{subsec.FEMCEM}.
Here, the conventional locally supported piecewise polynomial FE-bases 
$\{\phi_i(\vec r)\}$ and $\{\psi_j(\vec r)\}$
for $\sigma$ and $u$ are used.

Denote the eigenvalues and eigenvectors of $\Gamma_{\sigma}$
by $\lambda^{\sigma}_i$ and $v^{\sigma}_i$, respectively,
and those of $\Gamma_u$ by 
$\lambda^{u}_i$ and $v^{u}_i$.
The POD bases
$\{\phi^{\mathrm{POD}}_i(\vec r)\}$
and
$\{\psi^{\mathrm{POD}}_j(\vec r)\}$
are
\begin{eqnarray}
\phi^{\mathrm{POD}}_i(\vec r) &=& \sum_{k=1}^N v^{\sigma}_i(k)\phi_k(\vec r),\ 
i=1,\ldots,\hat N\\
\psi^{\mathrm{POD}}_i(\vec r) &=& \sum_{k=1}^M v^u_i(k)\psi_k(\vec r),\ 
i=1,\ldots,\hat M
\end{eqnarray}
Due to the construction, 
we have for the reduced order conductivity parameters
$\alpha^{\mathrm{POD}} = 
(\alpha_1^{\mathrm{POD}},\ldots,\alpha_{\hat N}^{\mathrm{POD}})\trans$
in (\ref{eq.sigmaapprx})
\begin{eqnarray}
{\mathbb E}\{\alpha^{\mathrm{POD}}\} = [0,\ldots,0]\trans 
\label{eq.E_alpha}
\\
\Gamma_{\alpha^{\mathrm{POD}}} = 
\mathrm{diag}(\lambda^{\sigma}_1,\ldots,\lambda^{\sigma}_{\hat N})
\label{eq.Gamma_alpha}
\end{eqnarray}
where $\Gamma_{\alpha^{\mathrm{POD}}}$ denotes the 
covariance matrix of $\alpha^{\mathrm{POD}}$.

The reduced-order observation model of EIT is constructed
by inserting the approximations (\ref{eq.sigmaapprx}) and
(\ref{eq.uapprx}) into
the variational form 
(\ref{eq.variational.form}).
This yields, equivalently with (\ref{eq.atf}), a system
\be
A^{\mathrm{POD}} \theta^{\mathrm{POD}} = f.
\label{eq.atfPOD}
\ee
Here $\theta^{\mathrm{POD}} = 
\left[{\beta^{\mathrm{POD}}}\trans, \gamma\trans \right]\trans$,\
$\beta^{\mathrm{POD}} = 
[\beta^{\mathrm{POD}}_0,\ldots,\beta^{\mathrm{POD}}_{\hat M}]\trans$
and
\be
A^{\mathrm{POD}} = \left[ \begin{array}{cc} B^{\mathrm{POD}} 
+ D^{\mathrm{POD}} & E^{\mathrm{POD}} \mathcal{C} \\ 
\mathcal{C}\trans {E^{\mathrm{POD}}}\trans & \mathcal{C}\trans F \mathcal{C} \end{array} \right]
\ee
where
\begin{eqnarray}
B^{\mathrm{POD}} &=& \sum_{k=0}^{\hat N} \alpha_k^{\mathrm{POD}}
B_k^{\mathrm{POD}}
\label{eq.APOD0}\\
B_k^{\mathrm{POD}}(i,j) &=& \int_\Omega \phi_k^{\mathrm{POD}}\ \nabla \psi_i^{\mathrm{POD}} \cdot \nabla \psi_j^{\mathrm{POD}} \dd \Omega,\quad 0 \leq i,j \leq \hat{M} \label{eq.APOD1}\\
D^{\mathrm{POD}}(i,j) &=& \sum_{\ell=1}^L \frac{1}{z_\ell} \int_{e_\ell} \psi^{\mathrm{POD}}_i \psi^{\mathrm{POD}}_j \dd S,\quad 0 \leq i,j \leq \hat{M} \label{eq.APOD2}\\
E^{\mathrm{POD}}(i,j) &=& -\frac{1}{z_j} \int_{e_j} \psi^{\mathrm{POD}}_i \dd S,\quad 0\leq i \leq \hat{M},\ 0\leq j \leq L \label{eq.APOD3}
\end{eqnarray}
and $f, \gamma, z_j, \mathcal{C}$ and $F$ are defined in Section \ref{subsec.FEMCEM}.
Here, for notational convenience,
the expectations of $\sigma$ and $u$ are denoted by
$\phi^{\mathrm{POD}}_0$ and $\psi^{\mathrm{POD}}_0$.
Hence,
$\phi^{\mathrm{POD}}_0(\vec r) = \sigma_0,\ 
\alpha^{\mathrm{POD}}_0 = 1,\ 
\psi^{\mathrm{POD}}_0(\vec r)=u_0$ and
$\beta^{\mathrm{POD}}_0 = 1$.
When solving the system (\ref{eq.atfPOD}),
the terms depending on $\beta^{\mathrm{POD}}_0$ are moved to the
right hand side of the equation.

Because the POD bases used in the reduced-order approximation of $u$
are globally 
supported, matrix $A^\mathrm{POD}$ is dense.
However, 
if the number of selected POD modes ($\hat M$) is small,
the dimension of $A^\mathrm{POD}$ is low, making the solution of the 
reduced-order CEM computationally inexpensive.
The same applies to computing the Jacobian matrix of 
the reduced-order observation mapping.

We complete the reduced-order observation model
by accounting for the error caused by the model reduction.
Here, we adopt the approximation error method
\cite{kaipio2005}
which is based on statistical modelling of errors.
Previously,
the approximation error method has been used 
for recovering from errors caused e.g. by
discretization \cite{kaipio2005,nissinen2009},
uncertainty of the geometry \cite{heino2005},
unknown boundary data \cite{lehikoinen2007}
and unknown contact impedances 
\cite{nissinen2008,nissinen2009}.
Here, we rewrite the observation equation (\ref{obs.eq}) in the form
\be\label{obs.eq.POD}
V_{\epsilon} = \mathcal{V}^{\mathrm{POD}}(\alpha^{\mathrm{POD}}) + \epsilon''
\ee
where 
$\mathcal{V}^{\mathrm{POD}}(\alpha^{\mathrm{POD}})$ denotes the 
reduced-order counterpart of the mapping  $\mathcal{V}(\sigma)$ in
(\ref{obs.eq}).
Further, the error term 
$\epsilon''$ is of the form
$\epsilon'' = \epsilon + \epsilon'$ 
where $\epsilon$ is the measurement noise, and
\be\label{eq.espilon}
\epsilon' = \mathcal{V}(\sigma) - \mathcal{V}^{\mathrm{POD}}(\alpha^{\mathrm{POD}})
\ee
is the error caused by the reduced-order approximations.
To construct a model for the statistics of the error term $\epsilon'$,
we compute
$\epsilon'_{(i)} = \mathcal{V}(\sigma^{(i)}) - 
\mathcal{V}^{\mathrm{POD}}(\alpha_{(i)}^{\mathrm{POD}})$
corresponding to all conductivity samples $\sigma^{(i)}$
in the sample set $\left\{\sigma^{(i)}\right\}_{i=1}^T$.
Here $\alpha_{(i)}^{\mathrm{POD}}$,
a sample of the reduced order representation of conductivity, is computed
by projecting $\sigma_{(i)}$ to the reduced order subspace spanned by
$\{\phi^{\mathrm{POD}}_k\}_{k=1}^{\hat N}$.
We approximate the expectation and the covariance matrix of $\epsilon'$ 
by the sample mean and covariance, and write a Gaussian approximation for
$\epsilon'$, such that
$\epsilon' \sim \mathcal{N}(\bar{\epsilon}', \Gamma_{\epsilon'})$.
Further, if $\epsilon'$ and the measurement noise $\epsilon$ are mutually uncorrelated,
we can write 
$\epsilon'' \sim \mathcal{N}(\bar{\epsilon}'', \Gamma_{\epsilon''})$.
where the expectation and the covariance of 
$\epsilon''$ are of the form 
$\bar{\epsilon}'' = \bar{\epsilon}+\bar{\epsilon}'$
and $\Gamma_{\epsilon''} = \Gamma_{\epsilon} + \Gamma_{\epsilon'}$,
respectively.

Using the reduced-order models,
we write the MAP estimate corresponding to (\ref{eq.mapmin})
in the form 
\begin{eqnarray}
\alpha^{\mathrm{POD}}_{\rm MAP} &=& 
\arg \min_{\alpha^{\mathrm{POD}}}
\left\{\vectornorm{L_{\alpha^{\mathrm{POD}}} 
\alpha^{\mathrm{POD}}}^2
+ 
\right. \nonumber \\
&& \left. 
\vectornorm{L_{\epsilon''} (V_{\epsilon}^\mathrm{meas} 
- \mathcal{V}^{\mathrm{POD}}(\alpha^{\mathrm{POD}}) - \bar{\epsilon}'')}^2 \right\} 
\label{eq.mapminPOD}
\end{eqnarray}
where we have noted that
according to equation (\ref{eq.E_alpha}) the expectation of the parameter vector
${\mathbb E}\{\alpha^{\mathrm{POD}}\} = [0,\ldots,0]\trans$.
Further,
$L_{\alpha^{\mathrm{POD}}}$ and $L_{\epsilon''}$ 
are the Cholesky factors of 
the inverted covariance matrices $\Gamma_{\alpha^{\mathrm{POD}}}^{-1}$
and $\Gamma_{\epsilon''}^{-1}$, respectively.
The form (\ref{eq.mapminPOD}) implies that $\epsilon''$ and $\alpha^{\mathrm{POD}}$
are mutually uncorrelated;
this approximation 
has turned out to be adequate in many cases \cite{kaipio2005}.
For modeling the cross-correlation of 
$\epsilon''$ and $\alpha^{\mathrm{POD}}$,
see \cite{kolehmainen2010}.

Finally, it should be noted that constructing matrix $A^\mathrm{POD}$
in the reduced order EIT forward model
is a time consuming task, especially because
it necessitates the sample set
$\left\{u^{(i)}\right\}_{i=1}^T$.
This set is constructed by
solving the (original, high dimensional)
forward problem of EIT $T$ times, 
and the number of samples ($T$) must be set large 
in order to retrieve the statistics of $u$ sufficiently.
However, constructing the sample set
does {\it not} require EIT measurement data;
it only depends on the measurement setup, geometry and the prior models.
The same applies to 
computing
the eigenvalue decompositions,
matrices 
$B_k^{\mathrm{POD}}$, $D^{\mathrm{POD}}$ and $E^{\mathrm{POD}}$,
and the approximation error statistics. 
Hence, all these time consuming tasks can be performed off-line,
before starting the measurements.
Once these precomputations are carried out,
the solution of the minimization problem (\ref{eq.mapminPOD})
is obtained with an iteration with reduced-order models;
each iteration step
necessitates only calculating the sum 
(\ref{eq.APOD0}), 
solving the low dimensional system (\ref{eq.atfPOD}),
(in Gauss-Newton method) computing the Jacobian matrix of the
reduced order mapping $\mathcal{V}^{\mathrm{POD}}$,
and computing the estimate update by solving a linear system
with $\hat N$ unknowns.
If
$\hat N << N$ and $\hat M << M$,
these tasks 
are significantly less demanding than those needed for solving the
original MAP-estimate (\ref{eq.mapmin}).

\vspace{-6pt}


\section{SIMULATIONS AND EXPERIMENTS} \label{sect.examples}
\vspace{-2pt}

In this section, the EIT reconstruction with the proposed 
POD based reduced-order model is evaluated with 
numerical and experimental tests.
The results are compared with the reconstructions obtained using
a conventional FE approximation for the CEM.
All the computations are carried out in Matlab environment with Dell Precision T7400
workstation (two quad core Intel Xeon E5420 CPUs and 32 Gb of RAM).


\subsection{Prior models and POD bases}
\label{subsec.prior.POD}

\begin{figure}
\centering\includegraphics[width=0.35\textwidth]{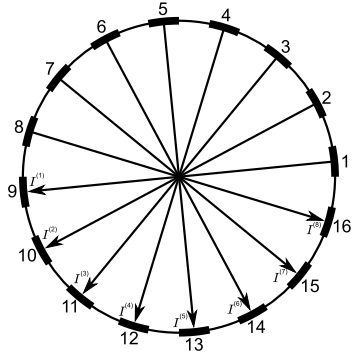}
\caption{Schematic figure of the electrode locations and numbering, 
and current injections $I^{(1)},\ldots,I^{(8)}$.}
\label{fig.electrodesandinj}
\end{figure}

We modeled the target domain as a circle with the diameter 28 cm.
In the model, 16 electrodes (width 2.5 cm) were set equidistantly on the boundary.
Electric currents with an amplitude of 1 mA were injected to the target with opposite
current injection scheme, see Figure \ref{fig.electrodesandinj}.
The contact impedances $z_\ell$ were set to
$0.01$ $\Omega\mathrm{cm}^2$ for all electrodes.
The 2D computational models described in this section
were used in 
the EIT reconstructions 
both in numerical simulation studies 
and in the experimental studies.
In the real data measurements described in 
Section \ref{subsec.realdata}, the target domain was a 3D cylinder 
(diameter of 28 cm, height 7 cm), the conductivity was
homogeneous in vertical direction and the electrodes extended from bottom to top
of the cylinder.
In such a case, the 2D approximation of the EIT model is usually adequate.
In all the inverse computations (corresponding to both conventional and
reduced-order models) we used FE meshes consisting of 2414 elements.
In the conventional FE scheme, we approximated the conductivity distribution in piecewise
linear basis and the potential distribution in piecewise quadratic basis.
The dimensions of the bases
$\{\phi_i(\vec r)\}_{i=1}^N$ 
and $\{\psi_j(\vec r)\}_{j=1}^M$
in the conventional FE-approximation of the CEM were 
$N=1263$
and 
$M=4939$.

\begin{figure}
\centering
\includegraphics[width=0.95\textwidth]{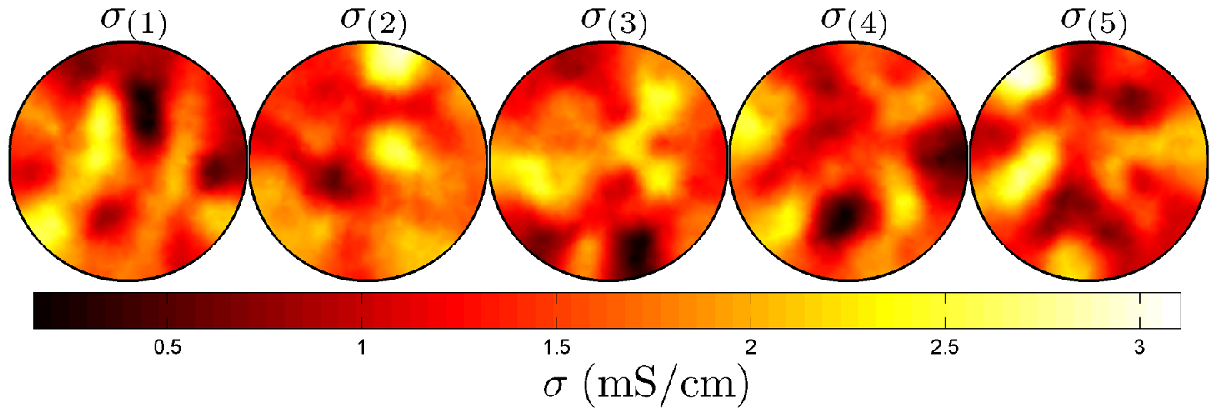}
\includegraphics[width=0.95\textwidth]{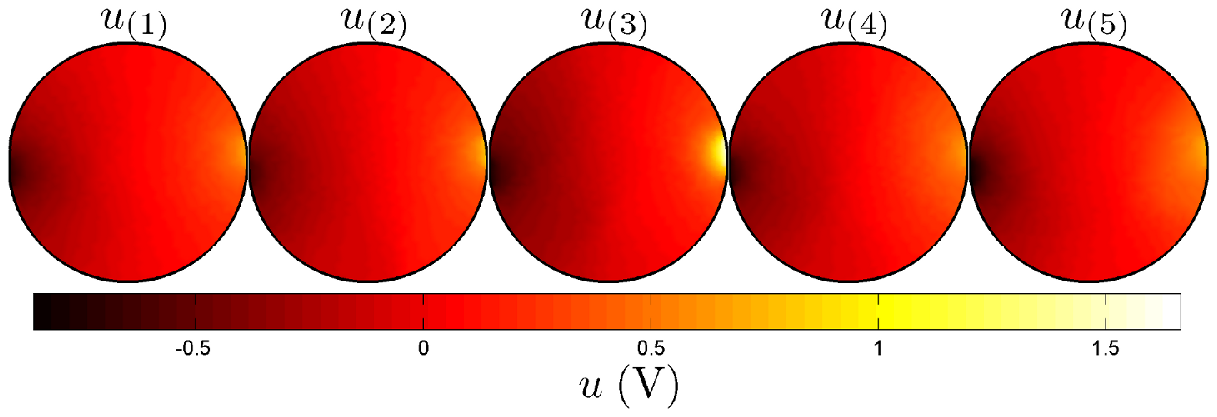}
\caption{
First row:
Conductivity samples 
$\sigma^{(i)},\ i=1,\ldots,5$ 
drawn from the
prior $\pi_\mathrm{pr,1}$.
Second row: Potentials $u^{(i)}$ corresponding to the conductivity samples
in upper figure.
All potential distributions correspond to the first current injection,
(cf. Fig. \ref{fig.electrodesandinj}).
\label{fig.samples1}}
\end{figure}

\begin{figure}
\centering
\includegraphics[width=0.95\textwidth]{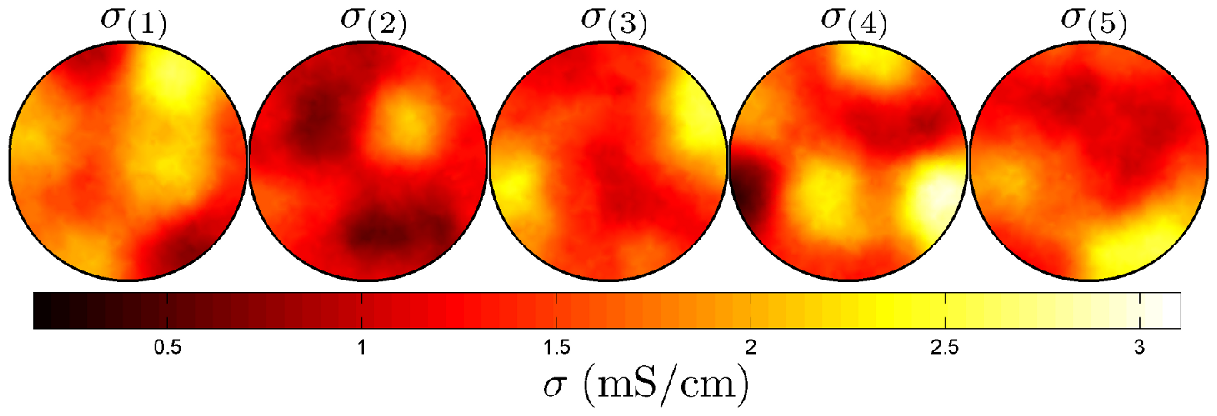}
\includegraphics[width=0.95\textwidth]{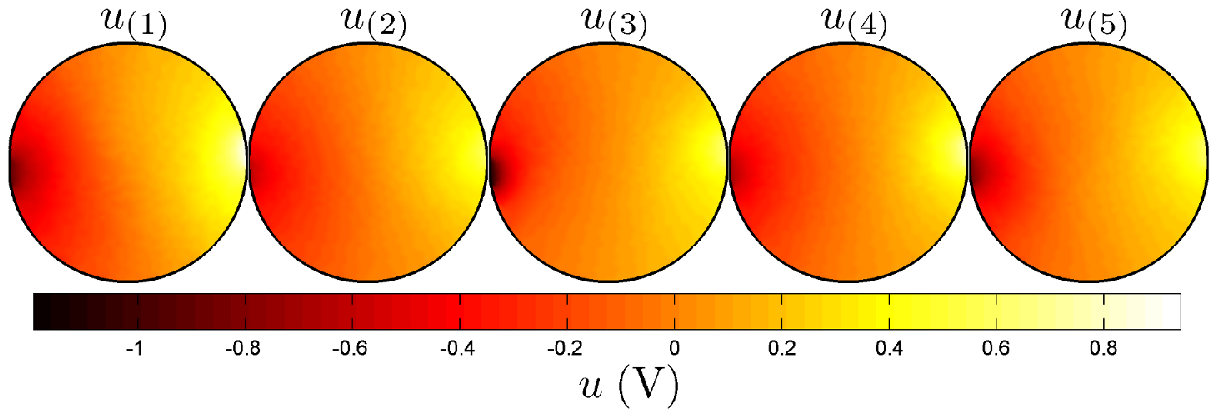}
\caption{
First row:
Conductivity samples 
$\sigma^{(i)},\ i=1,\ldots,5$ 
drawn from the
prior $\pi_\mathrm{pr,2}$.
Second row: Potentials $u^{(i)}$ corresponding to the conductivity samples
in upper figure.
All potential distributions correspond to the first current injection.
\label{fig.samples2}}
\end{figure}

In order to study the effect of the prior model to the model reduction,
we constructed the POD based reduced-order models corresponding to two
different prior distributions.
We denote the probability densities corresponding to the two priors
by $\pi_\mathrm{pr,1}(\sigma)$ and $\pi_\mathrm{pr,2}(\sigma)$.
Both of the prior models were selected to be of the form of
the proper (informative) smoothness prior
\cite{calvetti2006,kaipio2005,kolehmainen2007a}.
In the two models, the variances of the
nodal conductivity values $\sigma_i=\sigma(\vec r_i)$ 
were equal,
$\mathrm{var}(\sigma_i)=0.25 \mathrm{\mu S}^{2}\mathrm{cm}^{-2},\ i=1,\ldots,N$.
The degree of the spatial smoothness, however, differed between the two models;
in the model $\pi_\mathrm{pr,1}(\sigma)$, the cross-covariance
$\mathrm{cov}(\sigma_i,\sigma_j)$ decreased faster with the 
distance between the coordinates $\vec r_i$ and $\vec r_j$
than in the model $\pi_\mathrm{pr,2}(\sigma)$.
Five random samples corresponding to prior models
$\pi_\mathrm{pr,1}(\sigma)$ and $\pi_\mathrm{pr,2}(\sigma)$
are depicted on top rows of
Figures \ref{fig.samples1} and \ref{fig.samples2}, 
respectively.
As expected,
the samples corresponding to $\pi_\mathrm{pr,2}(\sigma)$
are smoother than those corresponding to $\pi_\mathrm{pr,1}(\sigma)$.
On the bottom rows of
Figures \ref{fig.samples1} and \ref{fig.samples2},
the electric potential fields corresponding to the selected conductivity samples
and the first current injection are illustrated.

\begin{figure}
\centering
\includegraphics[width=0.95\textwidth]{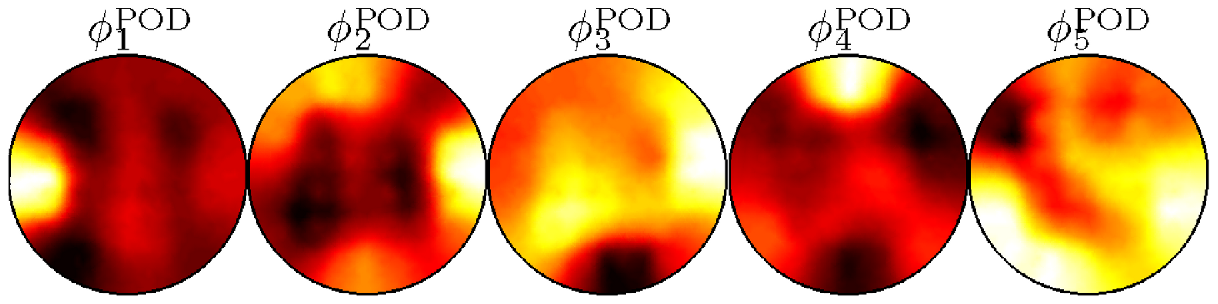}
\includegraphics[width=0.95\textwidth]{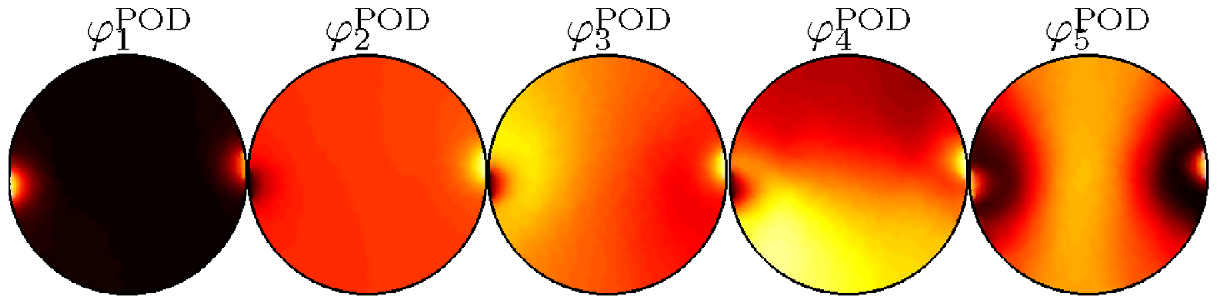}
\caption{
POD bases corresponding to prior density $\pi_\mathrm{pr,1}$.
First row: POD bases $\phi^{\mathrm{POD}}_i,\ i=1,\ldots,5$ 
of the conductivity.
Second row: POD bases $\varphi^{\mathrm{POD}}_{i},\ i=1,\ldots,5$ of the potential.
\label{fig.basis1}}
\end{figure}

\begin{figure}
\centering
\includegraphics[width=0.95\textwidth]{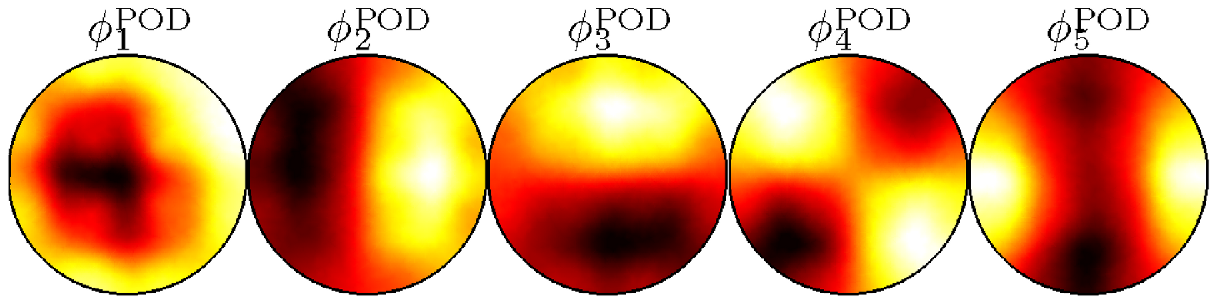}
\includegraphics[width=0.95\textwidth]{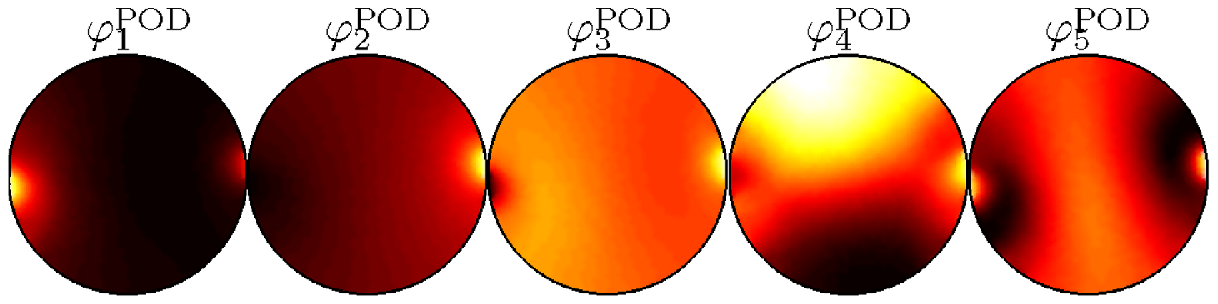}
\caption{
POD bases corresponding to prior density $\pi_\mathrm{pr,2}$.
First row: POD bases $\phi^{\mathrm{POD}}_i,\ i=1,\ldots,5$ 
of the conductivity.
Second row: POD bases $\varphi^{\mathrm{POD}}_{i},\ i=1,\ldots,5$ of the potential.
\label{fig.basis2}}
\end{figure}

The POD modes corresponding to the priors $\pi_\mathrm{pr,1}$ and $\pi_\mathrm{pr,2}$
were constructed as described in Section \ref{sect.pod}.
The sample sets consisted of 7500 
conductivity and potential samples.
In the selected test case, 
the current injections and prior distribution for the conductivity
were rotationally symmetric.
Hence, the POD basis for the potential was only generated 
corresponding to the first current injection $I^{(1)}$;
the potential bases corresponding the other current injections
$I^{(i)},\ i=2,\ldots,8$
were constructed by rotating the first basis by $360/16\cdot (i-1)$
degrees.
If the current injections or prior distribution were not rotationally symmetric,
the POD basis would have been generated for each current injection separately.
The first five POD bases of the conductivity
and the potential corresponding to the prior models 
$\pi_\mathrm{pr,1}$ and $\pi_\mathrm{pr,2}$
are drawn in Figures \ref{fig.basis1} and \ref{fig.basis2}, respectively.
As expected, the POD bases $\phi^{\mathrm{POD}}_i$ 
of the conductivity $\sigma$ are spatially smoother in the case
of prior $\pi_\mathrm{pr,2}$
than in the case of prior $\pi_\mathrm{pr,1}$.
The POD bases $\varphi_{i}$ of the potential $u$
possess qualitatively similar features in cases of
the two different prior models.

\begin{figure}
\centering
\includegraphics[width=0.45\textwidth]{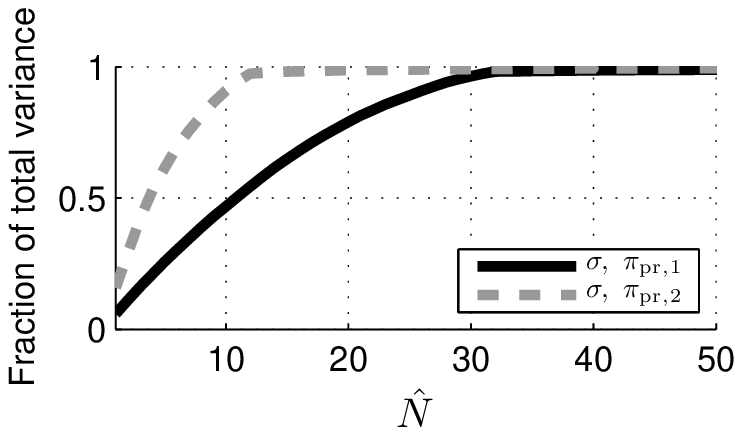}
\hspace{0.05\textwidth}
\includegraphics[width=0.45\textwidth]{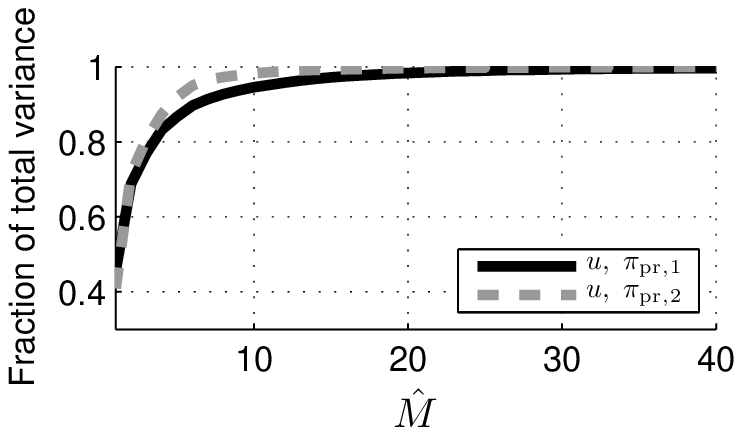}
\caption{Left: $\chi_{\sigma}$,
the fraction of the variance retained by the reduced-order representation of 
$\sigma$ as function of $\hat N$, the dimension of the reduced order
basis $\{\phi^{\mathrm{POD}}_i\}$.
Right: $\chi_u$,
the fraction of the variance retained by the reduced-order representation of 
$u$ as function of $\hat M$, the dimension of the reduced order
basis $\{\psi^{\mathrm{POD}}_j\}$.
The the solid black
lines correspond to the prior model $\pi_\mathrm{pr,1}$
and the dashed gray lines to the prior model $\pi_\mathrm{pr,2}$
\label{fig.eigs}}
\end{figure}

The fractions of the variances that are captured by using
$\hat N$- and $\hat M$-order approximations
of $\sigma$ and $u$, respectively,
are given by
$\chi_{\sigma}(\hat N) 
=(\sum_{k=1}^N\lambda^{\sigma}_k)^{-1}
\sum_{\ell=1}^{\hat N}\lambda^{\sigma}_{\ell}$
and
$\chi_{u}(\hat M) 
= (\sum_{k=1}^M\lambda^{u}_k)^{-1}
\sum_{\ell=1}^{\hat M}\lambda^{u}_{\ell}$.
Figure \ref{fig.eigs} displays
$\chi_{\sigma}(\hat N)$ and $\chi_{u}(\hat M)$ for both
prior models $\pi_\mathrm{pr,1}$ and $\pi_\mathrm{pr,2}$.
A rapid convergence of the retained variance fraction 
to 1 indicates 
that an accurate POD approximation requires only a small number of 
POD bases.
Clearly, in the case of prior $\pi_\mathrm{pr,2}$,
the retained variance fraction $\chi_{\sigma}$ increases more
rapidly than in the case of prior $\pi_\mathrm{pr,1}$.
This result is intuitively appealing:
when a random field features a very high spatial smoothness
(prior model $\pi_\mathrm{pr,2}$),
its variations can be represented in a low dimensional
basis.
For the same reason, 
$\chi_{u}$ increases more rapidly than $\chi_{\sigma}$ --
indeed,
the samples of the electric potential $u$ are spatially much smoother than the samples
of the conductivity $\sigma$ (cf. Figs. \ref{fig.samples1} and \ref{fig.samples2}).
Further,
the  properties of $\sigma$ are reflected by 
the smoothness of $u$ via the diffusion model (\ref{eq.CEM1});
the smoother $\sigma$ is, the smoother is also $u$.
In consequence,
$\chi_{u}$ increases more rapidly
in the case of  prior $\pi_\mathrm{pr,2}$
than in the case of prior $\pi_\mathrm{pr,1}$.

In this section,
we demonstrated the effect of the prior model to the
POD based model reduction
by illustrating the samples, POD bases and retained variance fractions
in cases of to two prior models 
$\pi_\mathrm{pr,1}(\sigma)$ and $\pi_\mathrm{pr,2}(\sigma)$
corresponding to different degrees or spatial smoothness.
Below,
we evaluate the POD based EIT reconstruction.
The MAP estimates for the conductivity distributions
are computed only
using the prior model $\pi_\mathrm{pr,1}(\sigma)$,
that assumes higher spatial variation for the conductivity.


\subsection{Simulation results}

\begin{figure}
\centering\includegraphics[width=0.7\textwidth]{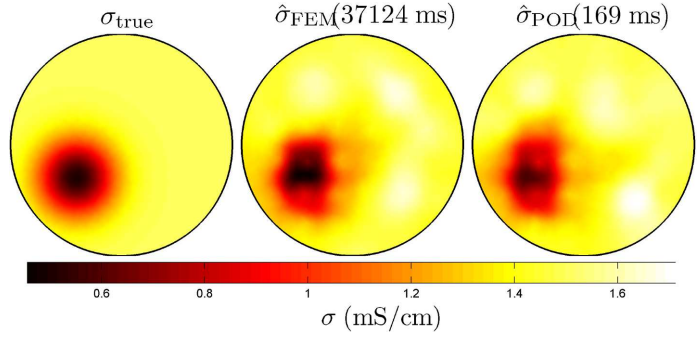}
\caption{Test case 1: Left: True simulated target $\sigma_\mathrm{true}$. 
Middle: Conventional reconstruction
$\hat\sigma_\mathrm{FEM}$. Right: Reduced-order reconstruction $\hat\sigma_\mathrm{POD}$. 
The reconstruction times
are shown in the parenthesis on top of the reconstructions.}
\label{fig.reconstruction1}
\end{figure}

\begin{figure}
\centering\includegraphics[width=0.7\textwidth]{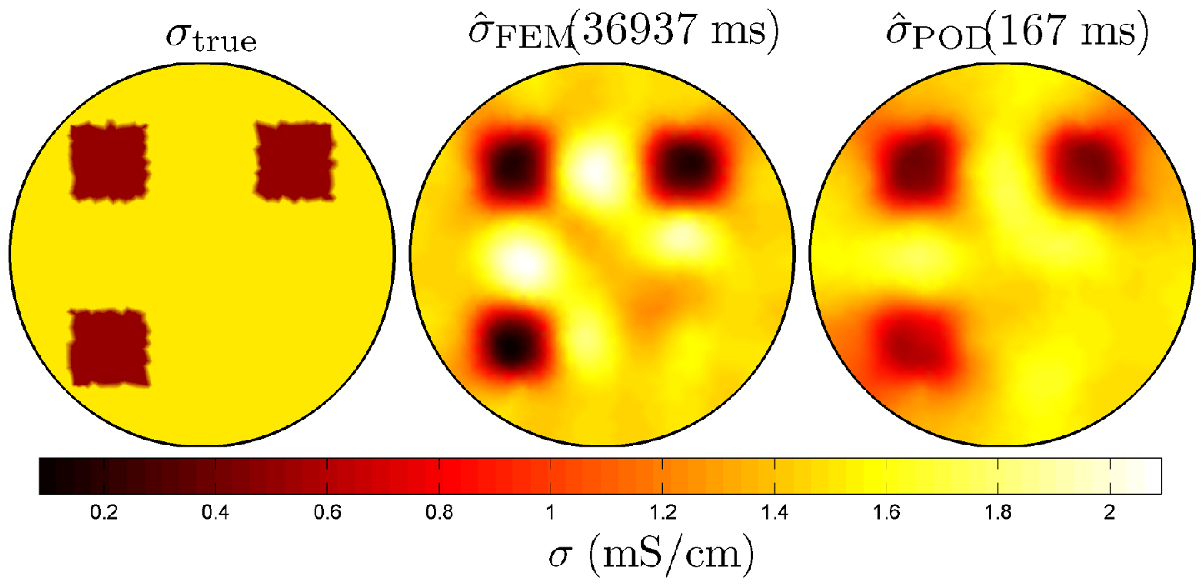}
\caption{Test case 2: Left: True simulated target $\sigma_\mathrm{true}$. 
Middle: Conventional reconstruction
$\hat\sigma_\mathrm{FEM}$. Right: Reduced-order reconstruction $\hat\sigma_\mathrm{POD}$. 
The reconstruction times
are shown in the parenthesis on top of the reconstructions.}
\label{fig.reconstruction2}
\end{figure}

\begin{figure}
\centering\includegraphics[width=0.7\textwidth]{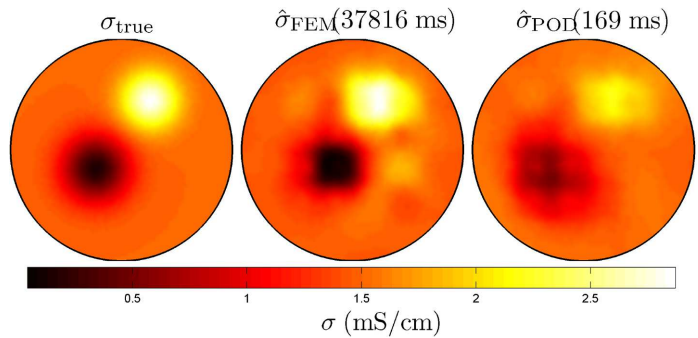}
\caption{Test case 3: Left: True simulated target $\sigma_\mathrm{true}$. 
Middle: Conventional reconstruction
$\hat\sigma_\mathrm{FEM}$. Right: Reduced-order reconstruction $\hat\sigma_\mathrm{POD}$. 
The reconstruction times
are shown in the parenthesis on top of the reconstructions.}
\label{fig.reconstruction3}
\end{figure}

The simulation tests were carried out with three different 
target distributions:
a target with a smooth resistive inclusion (test case 1),
a target with three rectangular resistive inclusions with sharp boundaries (test case 2), 
and a target with both
a resistive and a conductive inclusion (test case 3).
The targets are illustrated in 
Figures \ref{fig.reconstruction1}-\ref{fig.reconstruction3}
(left).

The simulated measurement data were computed using 
the conventional FE approximation of the measurement model.
The FE mesh consisted of 8394 elements.
The conductivity distribution was approximated in piecewise
linear basis and the potential distribution in piecewise quadratic basis.
The number of nodes in the 1st order FE mesh was 4374
and in the 2nd order mesh 17141.
Gaussian distributed noise was added to the simulated measurements.
The noise consists of two components: both of the components were of zero mean; 
the standard deviation (std) of the first
component was 1\% of the absolute value of the noiseless voltages, 
and the std of the second component was 0.1\% of the
difference between the maximum and minimum voltages.

The EIT reconstructions with both the
conventional FE approximations and the POD based model reduction
were computed.
In the former case, the reconstruction
was computed by solving the minimization problem
(\ref{eq.mapmin}),
and in the latter case
by solving
(\ref{eq.mapminPOD}).
Both optimization problems were solved with the
Gauss-Newton method, employed with a line-search
\cite{vauhkonen2004}.
The same prior model 
$\pi_\mathrm{pr}(\sigma)=\pi_\mathrm{pr,1}(\sigma)$
(see Section \ref{subsec.prior.POD})
was used both in conventional and in the
reduced-order reconstructions.
Further, in both cases, the variance of the measurement noise
was assumed to be known.
The numbers of POD modes both for  $\sigma$ and $u$ were selected such that $99 \%$
of the variances of the respective random variables were
retained.
This yielded reduced-order representations consisting of 54 basis vectors for the 
conductivity
and 25 basis vectors for the potential.

The reconstructions corresponding 
to both the conventional and the reduced-order model are
depicted in Figures \ref{fig.reconstruction1}-\ref{fig.reconstruction3}
(middle and right).
The times required for each reconstruction are also 
indicated in the figures.
In test case 1, 
the position of the inclusion was well tracked in both reconstructions.
The computation time for the reconstruction corresponding 
to the conventional model was about 37 s,
and for the reconstruction with the reduced-order model 
significantly less, below 200 milliseconds.
That is, the reconstruction times were more than two orders of magnitude 
smaller in the case of the reduced-order model.
In test cases 2 and 3, the reconstruction times were approximately the same as in test case 1.
Also in these test cases, both standard and reduced-order 
reconstructions correspond well to the true targets.
In all test cases, 
the POD-based reconstructions are only slighly smoother than
those based on standard FE bases.

\begin{figure}
\centering\includegraphics[width=0.7\textwidth]{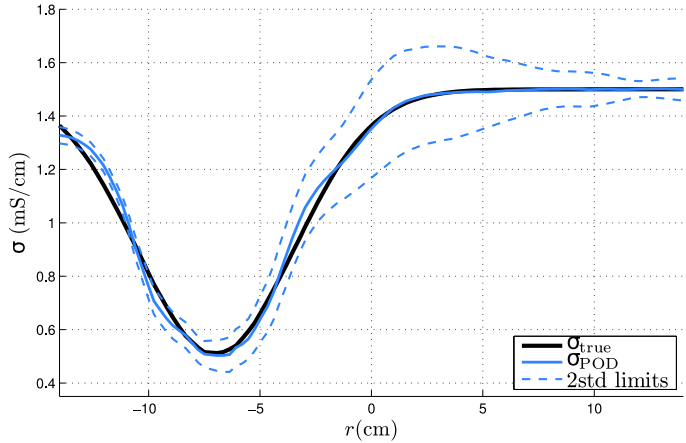}
\caption{Test case 1: Profiles of the true conductivity distribution 
(solid black line), the estimated conductivity $\hat\sigma_\mathrm{POD}$
(solid blue line) and the 
2 std limits of the estimate
(dashed blue lines).}
\label{fig.reconstruction4}
\end{figure}

In test case 1, the reliability of the 
POD based
reconstruction was assessed 
by computing the confidence 
limits defined by
two standard deviations (std) for the 
reconstructed conductivity.
The profiles of the true and the reconstructed conductivities, 
and the 2 std limits on a cross-section of the target domain
are plotted in Figure \ref{fig.reconstruction4}.
Here, we selected the cross-section such that
it diagonally passes through the midpoints of the inclusion and the
center of the circular target domain.
The figure shows that the true conductivity is mostly between 
the 2 std limits.
The figure also reveals that the 
uncertainty of the estimate is lowest near the
target boundaries and the highest
in the middle of the target.
This is because the sensitivity of EIT measurements
is highest near the target boundary where the electrodes are attached.

\begin{figure}
\centering\includegraphics[width=0.9\textwidth]{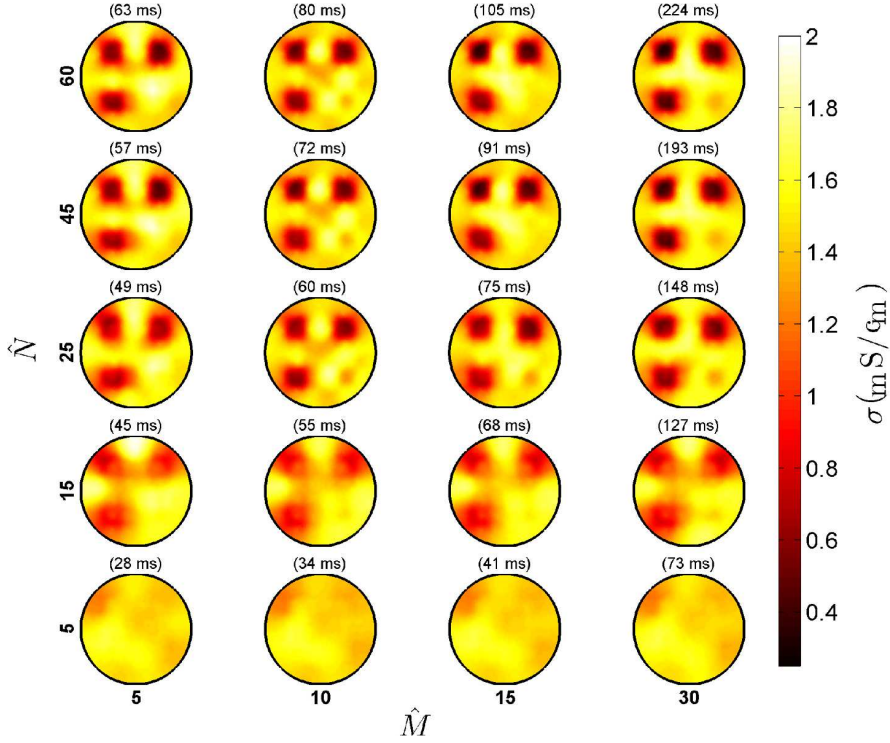}
\caption{Reduced-order reconstructions and reconstruction times with different
numbers of basis functions for $\sigma$ ($\hat N$) and $u$ ($\hat M$). 
The true target is shown
in Figure \ref{fig.reconstruction3}.}
\label{fig.numbasis}
\end{figure}

To study the effects of dimensions $\hat N$ and $\hat M$
of the POD bases
$\{\phi^{\mathrm{POD}}_i\}_{i=1}^{\hat N}$ and 
$\{\psi^{\mathrm{POD}}_j\}_{j=1}^{\hat M}$ 
to reduced-order reconstructions,
we computed the MAP estimates 
(\ref{eq.mapminPOD})
corresponding to test case 2
using various choices of $\hat N$ and $\hat M$.
In Figure \ref{fig.numbasis},
the results are presented in the form of a table 
where $\hat N$ increases from bottom to top,
and $\hat M$ increases from left to right.
The computation times are written
above the corresponding reconstructions.
The results demonstrate that especially 
the number of basis functions for the potential ($\hat M$)
can be set very small without ruining the quality
of the reconstructions -- even the reconstructions with 
$\hat M=5$ are relatively good, given that 
$\hat N$ is large enough ($\hat N \geq 45$).
Clearly, the 
number of basis functions required for 
approximation the conductivity ($\hat N$)
is higher than that of potential
-- the reconstructions corresponding to $\hat N \leq 15$
are severely blurred.
These are expected results,
because with the selected prior model
$\pi_\mathrm{pr}(\sigma)=\pi_\mathrm{pr,1}(\sigma)$
the number of basis functions required for retaining essentially all variations of
$\sigma$ is $\hat N\approx 30$,
while retaining the variations of $u$ requires somewhat less basis functions,
cf. Figure \ref{fig.eigs}.

\subsection{Experimental results}
\label{subsec.realdata}

\begin{figure}
\centering\includegraphics[width=0.8\textwidth]{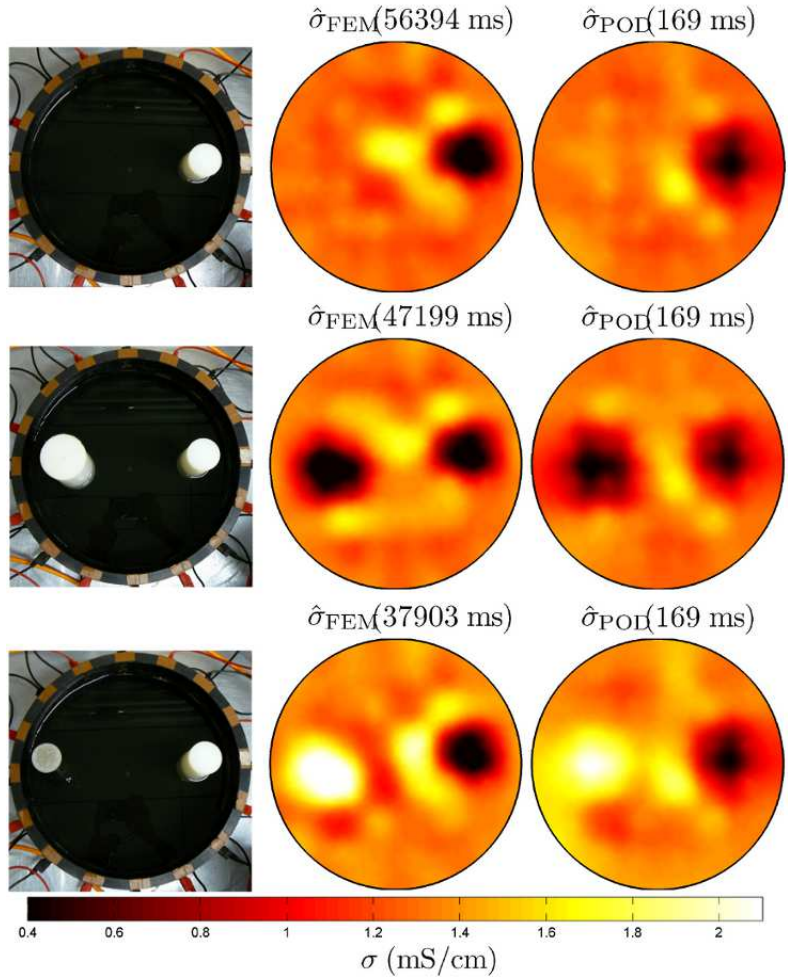}
\caption{Experimental tests: Left column: True targets. The brown markings on the boundary of the tank
indicate the electrode locations. 
Middle column: Conventional reconstructions $\hat\sigma_{\mathrm{FEM}}$.
Right column: Reduced-order reconstructions $\hat\sigma_{\mathrm{POD}}$.
The reconstruction times are shown in the parenthesis on top of the reconstructions.}
\label{fig.realdata}
\end{figure}

In the experiments, a cylindrical tank was filled with tap water, 
and three different target conductivities
were constructed by inserting both resistive and
conductive objects into the tank.
The first target included one  plastic (resistive) bar,
the second target two plastic bars,
and the third target contained one plastic and one metallic bar.
All these inclusions were homogeneous in the vertical direction.
Photos of all three targets are shown in the left column of 
Figure \ref{fig.realdata}.
The diameter of the tank was 28 cm and the height of the water level was 7 cm.
The electrode configuration in the tank consisted of 16 uniformly positioned, 25 mm 
wide boundary electrodes,
and they extended vertically from bottom to height of the water level.
In Figure \ref{fig.realdata},
the brown stripes on the tank wall indicate the locations of the electrodes.
The measurements were carried out using opposite current injection scheme as
in the simulations.
As the measurement system, the SIPFIN measurement device was used
\footnote{SIPFIN is a modification of the Radic Research SIP256 instrument,
see http://www.radic-research.de}.

Again, the EIT reconstructions
were computed both based on a conventional FE approximation of the CEM
and with the POD reduced-order model.
Same models were used 
as in the reconstructions based on simulated data in
the previous section.
The same numbers of POD basis functions
were selected as in simulations studies with targets 1-3,
i.e. $\hat N=54$ and $\hat M=25$.

The results are shown in Figure \ref{fig.realdata}.
In the reconstructions,
both corresponding to
the conventional FE approximation of the CEM
and the POD reduced-order model,
the inclusions are clearly visible and the positions of the 
inclusions correspond to the real targets.
In all test cases, 
the qualities of the reconstructions 
corresponding to the POD reduced-order model
are comparable to those corresponding to the
conventional FE based model.
Again, the computation times were remarkably shorter
when reduced-order model was used:
with the conventional FE approximation, the computation times
varied between 37903 ms and 56394 ms, while 
with the 
POD approximations the computations took only 169 ms in all test cases.

Some artifacts are present in all reconstuctions.
These are due to 
1) modeling errors that were not accounted for in the observation model.
Such modeling errors are the discretization error and the
error resulting from the unknown contact impedances.
These errors could be accounted for by the approximation error modeling
\cite{kaipio2005,nissinen2008,nissinen2009}.
2) Secondly,
the selected prior model, the smoothness prior,
is obviously not the best possible choice in the test cases.
Indeed, 
accross the boundaries of the inclusions,
the conductivity is discontinuous,
and the conductivity contrasts are high.
In such cases the assumption of smoothness is not a feasible one.
However,
as the aim of this study was only to demonstrate the
feasibility of the POD based model reduction,
we omit the considerations of other modeling errors
and prior models from this paper.

\vspace{-6pt}

\section{CONCLUSIONS} \label{sect.conclusions}
\vspace{-2pt}

In this paper, a reduced-order model for EIT has been proposed.
The model is based on the 
POD representations of the 
electric conductivity distribution and the
potential distribution.
The model reduction is considered in the Bayesian inversion framework,
and the POD bases are constructed based on 
{\it a priori} information on the conductivity.
In the reconstructions, the errors caused by the 
model reduction are treated
with the approximation error method.
The reduced-order measurement model has been tested both
simulated simulated and experimental data.
The results show that by expressing the conductivity and the potential with
reduced-order basis representation, it is possible to
obtain feasible reconstructions with a low computational effort.
In the selected test cases,
the speedup was more than two orders of magnitude.
In this paper, 
the POD based model reduction was tested with 2D examples.
However,
the proposed method can be directly implemented
to computational 3D model of CEM
in the presented form.

\ack
This study was supported by the Academy of Finland (projects 250215 and 140280) and
Finnish Doctoral Programme in Computational Sciences (FICS).

\section*{References}
\bibliographystyle{plain}
\bibliography{podeit}

\begin{thebibliography}{10}

\bibitem{banks2000}
H.T. Banks, M.L. Joyner, B.~Wincheski, and W.P. Winfree.
\newblock {Nondestructive evaluation using a reduced-order computational
  methodology}.
\newblock {\em Inverse Problems}, 16:929--945, 2000.

\bibitem{banks2003}
H.T. Banks and G.M. Kepler.
\newblock {Reduced order computational methods for electromagnetic material
  interrogation using pulsed signals and conductive reflecting interfaces}.
\newblock {\em Journal of Inverse and Ill-posed Problems}, 11:343--370, 2003.

\bibitem{barber1984}
D.C. Barber and B.H. Brown.
\newblock {Applied potential tomography}.
\newblock {\em J Phys E: Sci Instrum}, 17:723--733, 1984.

\bibitem{brunet2010}
P.~Brunet, R.~Clement, and C.~Bouvier.
\newblock {Monitoring soil water content and deficit using Electrical
  Resistivity Tomography (ERT) - A case study in the Cevennes area, France}.
\newblock {\em Journal of Hydrology}, 380(1-2):146--153, 2010.

\bibitem{calvetti2006}
D.~Calvetti, J.P. Kaipio, and E.~Somersalo.
\newblock {Aristotelian prior boundary conditions}.
\newblock {\em Int. J. Math. Comp. Sci.}, 1:63--81, 2006.

\bibitem{cao2006}
Y.~Cao, J.~Zhu, Z.~Luo, and I.M. Navon.
\newblock {Reduced-order modeling of the upper tropical pacific ocean model
  using proper orthogonal decomposition}.
\newblock {\em Computers \& Mathematics with Applications}, 52(8-9):1373--1386,
  2006.

\bibitem{cheng1989}
K.-S. Cheng, D.~Isaacson, J.C. Newell, and D.G. Gisser.
\newblock {Electrode models for electric current computed tomography}.
\newblock {\em IEEE Trans Biomed Eng}, 36:918--924, 1989.

\bibitem{deane1991}
AE~Deane, IG~Kevrekidis, GE~Karniadakis, and SA~Orszag.
\newblock {Low-dimensional models for complex geometry flows: Application to
  grooved channels and circular cylinders}.
\newblock {\em Phys. Fluids A}, 3:2337--2354, 1991.

\bibitem{draper2003}
B.A. Draper, K.~Baek, M.S. Bartlett, and J.R. Beveridge.
\newblock {Recognizing faces with PCA and ICA}.
\newblock {\em Computer vision and image understanding}, 91(1-2):115--137,
  2003.

\bibitem{frerichs2000}
I.~Frerichs.
\newblock {Electrical impedance tomography (EIT) in applications related to
  lung and ventilation: a review of experimental and clinical activities}.
\newblock {\em Physiological measurement}, 21(2):R1--R21, 2000.

\bibitem{gokturkler2008}
G.~G{\"o}kt{\"u}rkler, C.~Balkaya, and Z.~Erhan.
\newblock {Geophysical investigation of a landslide: The Altindag landslide
  site, Izmir (western Turkey)}.
\newblock {\em Journal of Applied Geophysics}, 65(2):84--96, 2008.

\bibitem{gutierrez2000}
J.A. Gutierrez, T.~Dyakowski, M.S. Beck, and R.A. Williams.
\newblock {Using electrical impedance tomography for controlling hydrocyclone
  underflow discharge}.
\newblock {\em Powder Technology}, 108(2-3):180--184, 2000.

\bibitem{heino2005}
J.~Heino, E.~Somersalo, and J.P. Kaipio.
\newblock {Compensation for geometric mismodelling by anisotropies in optical
  tomography}.
\newblock {\em Opt. Exp.}, 13:296--308, 2005.

\bibitem{holmes1998}
P~Holmes, JL~Lumley, and G~Berkooz.
\newblock {\em {Turbulence, coherent structures, dynamical systems and
  symmetry}}.
\newblock Cambridge University Press, 1998.

\bibitem{hosseini2010}
S.~Hosseini, D.~Patel, F.~Ein-Mozaffari, and M.~Mehrvar.
\newblock {Study of solid-liquid mixing in agitated tanks through electrical
  resistance tomography}.
\newblock {\em Chemical Engineering Science}, 65(4):1374--1384, 2010.

\bibitem{hotelling1933}
H.~Hotelling.
\newblock {Analysis of a complex of statistical variables into principal
  components}.
\newblock {\em J. Educ. Psychol.}, 24:417--441, 1933.

\bibitem{jin2008}
B.~Jin.
\newblock {Fast Bayesian approach for parameter estimation}.
\newblock {\em International Journal for Numerical Methods in Engineering},
  76:230--252, 2008.

\bibitem{jolliffe2002}
I.T. Jolliffe.
\newblock {\em {Principal Component Analysis}}.
\newblock Springer-Verlag, 2002.

\bibitem{kaipio2000}
J.P. Kaipio, V.~Kolehmainen, E.~Somersalo, and M.~Vauhkonen.
\newblock {Statistical inversion and Monte Carlo sampling methods in electrical
  impedance tomography}.
\newblock {\em Inverse Problems}, 16:1487--1522, 2000.

\bibitem{kaipio1999c}
J.P. Kaipio, V.~Kolehmainen, M.~Vauhkonen, and E.~Somersalo.
\newblock {Inverse problem with structural prior information}.
\newblock {\em Inverse Problems}, 15(3):713--729, 1999.

\bibitem{kaipio2005}
J.P. Kaipio and E.~Somersalo.
\newblock {\em {Statistical and Computational Inverse Problems}}.
\newblock Springer New York, 2005.

\bibitem{kaipio2007a}
J.P. Kaipio and E.~Somersalo.
\newblock {Statistical inverse problems: discretization, model reduction and
  inverse crimes}.
\newblock {\em Journal of Computational and Applied Mathematics},
  198(2):493--504, 2007.

\bibitem{karhunen1946}
K.~Karhunen.
\newblock {Uber Lineare Methoden in der Wahrscheinlichkeitsrechnung}.
\newblock {\em Annales Academiae Sciientiarum Fennicae, Series AI:
  Mathematica-Physica}, 37:3--79, 1946.

\bibitem{kirby1990}
M.~Kirby and L.~Sirovich.
\newblock {Application of the Karhunen-Loeve procedure for the characterization
  of human faces}.
\newblock {\em IEEE Transactions on Pattern Analysis and Machine Intelligence},
  12(1):103--108, 1990.

\bibitem{kolehmainen2007a}
V.~Kolehmainen, J.P. Kaipio, and H.R.B. Orlande.
\newblock {Reconstruction of thermal conductivity and heat capacity using a
  tomographic approach}.
\newblock {\em International journal of heat and mass transfer},
  50(25-26):5150--5160, 2007.

\bibitem{kolehmainen2010}
V.~Kolehmainen, T.~Tarvainen, S.R. Arridge, and J.P. Kaipio.
\newblock {Marginalization of uninteresting distributed parameters in inverse
  problems - application to optical tomography}.
\newblock {\em Int J Uncertainty Quantification}, 1:1--17, 2011.

\bibitem{lehikoinen2007}
A.~Lehikoinen, S.~Finsterle, A.~Voutilainen, L.M. Heikkinen, M.~Vauhkonen, and
  J.P. Kaipio.
\newblock {Approximation errors and truncation of computational domains with
  application to geophysical tomography}.
\newblock {\em Inverse Problems and Imaging}, 1(2):371--389, 2007.

\bibitem{lieberman2010}
C.~Lieberman, K.~Willcox, and O.~Ghattas.
\newblock {Parameter and state model reduction for large-scale statistical
  inverse problems}.
\newblock {\em SIAM Journal on Scientific Computing}, 32:2523--2542, 2010.

\bibitem{loeve1946}
M.~L{\`o}eve.
\newblock {Fonctions al{\'e}atoires de second ordre}.
\newblock {\em Rev. Sci.}, pages 195--206, 1946.

\bibitem{loeve1955}
M.~L{\`o}eve.
\newblock {\em {Probability Theory}}.
\newblock Princeton, N.J.: Van Nostrand, 1955.

\bibitem{ma2002}
X.~Ma and G.~Karniadakis.
\newblock {A low-dimensional model for simulating three-dimensional cylinder
  flow}.
\newblock {\em J. Fluid Mech.}, 458:181--190, 2002.

\bibitem{nissinen2008}
A.~Nissinen, L.M. Heikkinen, and J.P. Kaipio.
\newblock {The Bayesian approximation error approach for electrical impedance
  tomography experimental results}.
\newblock {\em Meas. Sci. Technol.}, 19:015501 (9pp), 2008.

\bibitem{nissinen2009}
A.~Nissinen, L.M. Heikkinen, V.~Kolehmainen, and J.P. Kaipio.
\newblock {Compensation of errors due to discretization, domain truncation and
  unknown contact impedances in electrical impedance tomography}.
\newblock {\em Meas. Sci. Technol.}, 20:105504 (13pp), 2009.

\bibitem{pakzad2008}
L.~Pakzad, F.~Ein-Mozaffari, and P.~Chan.
\newblock {Using electrical resistance tomography and computational fluid
  dynamics modeling to study the formation of cavern in the mixing of
  pseudoplastic fluids possessing yield stress}.
\newblock {\em Chemical Engineering Science}, 63:2508--2522, 2008.

\bibitem{park2008}
B.-G. Park, J.-H. Moon, B.-S. Lee, and S.~Kim.
\newblock {An electrical resistance tomography technique for the monitoring of
  a radioactive waste separation process}.
\newblock {\em International Communications in Heat and Mass Transfer},
  35(10):1307--1310, 2008.

\bibitem{perrone2004}
A.~Perrone, A.~Ianuzzi, V.~Lapenna, P.~Lorenzo, S.~Piscitelli, E.~Rizzo, and
  F.~Sdao.
\newblock {High-resolution electrical imaging of the Carco d'Izzo earthflow
  (southern Italy)}.
\newblock {\em Journal of Applied Geophysics}, 56(1):17--29, 2004.

\bibitem{pursiainen2006b}
S.~Pursiainen.
\newblock {Two-stage reconstruction of a circular anomaly in electrical
  impedance tomography}.
\newblock {\em Inverse Problems}, 22:1689--1703, 2006.

\bibitem{pursiainen2006a}
S.~Pursiainen and H.~Hakula.
\newblock {A high-order finite element method for electrical impedance
  tomography}.
\newblock In {\em Progress In Electromagnetics Research Symposium 2006,
  Cambridge, USA}, pages 260--264, 2006.

\bibitem{ravindran2000}
S.S. Ravindran.
\newblock {A reduced-order approach for optimal control of fluids using proper
  orthogonal decomposition}.
\newblock {\em International Journal for Numerical Methods in Fluids},
  34(5):425--448, 2000.

\bibitem{santosa1990}
F.~Santosa and M.~Vogelius.
\newblock {A backprojection algorithm for electrical impedance imaging}.
\newblock {\em SIAM J Appl Math}, 50:216--243, 1990.

\bibitem{sirovich1987b}
L.~Sirovich and M.~Kirby.
\newblock {Low-dimensional procedure for the characterization of human faces}.
\newblock {\em Journal of the Optical Society of America A}, 4(3):519--524,
  1987.

\bibitem{somersalo1992}
E.~Somersalo, M.~Cheney, and D.~Isaacson.
\newblock {Existence and uniqueness for electrode models for electric current
  computed tomography}.
\newblock {\em SIAM J Appl Math}, 52:1023--1040, 1992.

\bibitem{takriff2009}
M.S. Takriff, A.A. Hamzah, S.K. Kamarudin, and J.~Abdullah.
\newblock {Electrical resistance tomography investigation of gas dispersion in
  gas-liquid mixing in an agitated vessel}.
\newblock {\em Journal of Applied Sciences}, 9(17):3110--3115, 2009.

\bibitem{vauhkonen1997}
M.~Vauhkonen, J.P. Kaipio, E.~Somersalo, and P.A. Karjalainen.
\newblock {Electrical impedance tomography with basis constraints}.
\newblock {\em Inverse Problems}, 13:523--530, 1997.

\bibitem{vauhkonen1999a}
P.J. Vauhkonen, M.~Vauhkonen, T.~Savolainen, and J.P. Kaipio.
\newblock {Three-dimensional electrical impedance tomography based on the
  complete electrode model}.
\newblock {\em IEEE Transactions on Biomedical Engineering}, 46:1150--1160,
  1999.

\bibitem{vauhkonen2004}
P.J. Vauhkonen, M.~Vauhkonen, A.~Sepp\"{a}nen, and Kaipio J.P.
\newblock {Iterative image reconstruction in three-dimensional electrical
  impedance tomography}.
\newblock In {\em Proc. Inverse Problems, Design and Optimization Symposium,
  Rio de Janeiro, Brazil}, 2004.

\bibitem{vilhunen2002a}
T.~Vilhunen, J.P. Kaipio, P.J. Vauhkonen, T.~Savolainen, and M.~Vauhkonen.
\newblock {Simultaneous reconstruction of electrode contact impedances and
  internal electrical properties. Part I: Theory}.
\newblock {\em Meas. Sci. Technol.}, 13:1848--1854, 2002.

\bibitem{williams1995b}
R.A. Williams and M.S. Beck, editors.
\newblock {\em {Process Tomography, Principles, Techniques and Applications}}.
\newblock Oxford: Butterworth-Heinemann, 1995.

\bibitem{williams1997}
R.A. Williams, F.J. Dickin, J.A. Gutierrez, T.~Dyakowski, and M.S. Beck.
\newblock {Using electrical impedance tomography for controlling hydrocyclone
  underflow discharge}.
\newblock {\em Control Engineering Practice}, 5(2):253--256, 1997.

\bibitem{yeung2001}
K.Y. Yeung and W.L. Ruzzo.
\newblock {Principal component analysis for clustering gene expression data}.
\newblock {\em Bioinformatics}, 17(9):763--774, 2001.

\bibitem{zhang2009}
L.~Zhang, W.~Dong, D.~Zhang, and G.~Shi.
\newblock {Two-stage image denoising by principal component analysis with local
  pixel grouping}.
\newblock {\em Pattern Recognition}, 43:1531--1549, 2009.

\bibitem{zou2003}
Y.~Zou and Z.~Guo.
\newblock {A review of electrical impedance techniques for breast cancer
  detection}.
\newblock {\em Medical engineering \& physics}, 25(2):79--90, 2003.

\end{thebibliography}

\end{document}